\documentclass[epj]{webofc}
\usepackage[utf8]{inputenc}
\usepackage[varg]{txfonts}
\usepackage{booktabs}
\usepackage{xcolor}
\definecolor{darkred}{rgb}{0.4,0.0,0.0}
\definecolor{darkgreen}{rgb}{0.0,0.4,0.0}
\definecolor{darkblue}{rgb}{0.0,0.0,0.4}
\usepackage[bookmarks,linktocpage,colorlinks,
    linkcolor = darkred,
    urlcolor  = darkblue,
    citecolor = darkgreen]{hyperref}
\wocname{EPJ Web of Conferences}
\woctitle{Lattice2017}

\newcommand{\al}{\ensuremath{\alpha} }
\newcommand{\Ga}{\ensuremath{\Gamma} }
\newcommand{\de}{\ensuremath{\delta} }
\newcommand{\De}{\ensuremath{\Delta} }
\newcommand{\eps}{\ensuremath{\epsilon} }
\newcommand{\la}{\ensuremath{\lambda} }
\newcommand{\si}{\ensuremath{\sigma} }
\newcommand{\Si}{\ensuremath{\Sigma} }
\newcommand{\Mdag}{\ensuremath{M^{\dag}} }
\newcommand{\muhat}{\ensuremath{\hat\mu} }
\newcommand{\nuhat}{\ensuremath{\hat\nu} }
\newcommand{\psibar}{\ensuremath{\overline\psi} }
\newcommand{\lsim}{\ensuremath{\lesssim} }
\newcommand{\gsim}{\ensuremath{\gtrsim} }
\newcommand{\vev}[1]{\ensuremath{\left\langle #1 \right\rangle} }
\newcommand{\pf}{\ensuremath{\text{pf}\,} }
\newcommand{\nn}{\nonumber}
\newcommand{\eq}[1]{Eq.~(\ref{#1})}
\newcommand{\fig}[1]{Fig.~\ref{#1}}
\newcommand{\secref}[1]{Section~\ref{#1}}
\newcommand{\refcite}[1]{Ref.~\cite{#1}}

\begin{document}
\selectlanguage{english}
\title{Phases of a strongly coupled four-fermion theory}
\author{
  \firstname{David} \lastname{Schaich}\inst{1}\fnsep\thanks{Speaker, \email{schaich@itp.unibe.ch}} \and
  \firstname{Simon} \lastname{Catterall}\inst{2}
}
\institute{
  AEC Institute for Theoretical Physics, University of Bern, 3012 Bern, Switzerland \and
  Department of Physics, Syracuse University, Syracuse, New York 13244, United States
}

\abstract{
  We present ongoing investigations of a four-dimensional lattice field theory with four massless reduced staggered fermions coupled through an SU(4)-invariant four-fermion interaction.
  As in previous studies of four-fermion and Higgs--Yukawa models with different lattice fermion discretizations, we observe a strong-coupling phase in which the system develops a mass gap without breaking any lattice symmetry.
  This symmetric strong-coupling phase is separated from the symmetric weak-coupling phase by a narrow region of four-fermi coupling in which the system exhibits long-range correlations.
}

\maketitle

\section{Introduction} 
In this proceedings we present an update of our investigations into a simple four-dimensional lattice theory comprising four massless reduced staggered fermions coupled through an SU(4)-invariant four-fermion interaction~\cite{Catterall:2016dzf}.
Systems of this sort have received considerable interest in recent years within the condensed matter community~\cite{Fidkowski:2009dba, Morimoto:2015lua}, in the context of constructing models in which carefully chosen quartic interactions allow fermions to be gapped without breaking symmetries.
This feature of lattice four-fermion and Higgs--Yukawa models is predicted by strong-coupling arguments and was seen in earlier numerical calculations (including Refs.~\cite{Stephenson:1988td, Hasenfratz:1988vc, Lee:1989xq, Lee:1989mi, Bock:1990cx, Hasenfratz:1991it, Golterman:1992yha} and references therein).
These works generically found a three-phase structure, with a symmetric massless `paramagnetic weak-coupling' (PMW) phase separated from a symmetric massive `paramagnetic strong-coupling' (PMS) phase by a wide intermediate `ferromagnetic' (FM) phase characterized by a symmetry-breaking bilinear fermion condensate.
First-order transitions between these three phases left the symmetric massive PMS phase disconnected from the continuum limit. 

Recently, investigations of the three-dimensional version of the system we consider observed different behavior~\cite{Ayyar:2014eua, Ayyar:2015lrd, Catterall:2015zua, He:2016sbs}.
Three different numerical algorithms were used by these studies: fermion bags, rational hybrid Monte Carlo (RHMC) and quantum Monte Carlo.
Instead of the three phases described above, these works identified a direct transition between the massless analog of the PMW phase and a massive PMS-like phase at strong coupling.
These two phases appear to be separated by a continuous phase transition with non-Heisenberg exponents, raising the possibility of a new continuum limit at strong coupling.

Since the system we study possesses different exact lattice symmetries than those considered earlier, \textit{a priori} it is possible that this two-phase structure may persist in four dimensions.
The first work exploring the four-dimensional theory reported the reappearance of a broken FM phase~\cite{Ayyar:2016lxq, Ayyar:2016nqh}. 
However, in contrast to the earlier studies this intermediate phase was very narrow, and the transitions bounding it appeared consistent with second-order criticality with mean-field exponents.
These conclusions were not based on explicit measurements of bilinear condensates, but rather inferred from the volume scaling of a certain susceptibility.
Our own work in \refcite{Catterall:2016dzf} added source terms to the action in order to more directly address whether spontaneous symmetry breaking associated with the formation of specific bilinear condensates takes place.
We observed long-range correlations in the narrow critical region between the PMW and PMS phases, but did not observe spontaneous symmetry breaking and could not resolve whether this critical region corresponded to an intermediate phase or a single broad transition.

Since publishing \refcite{Catterall:2016dzf} we have discovered and corrected two factor-of-two errors in our publicly available code,\footnote{\texttt{\href{https://github.com/daschaich/fourfermion}{github.com/daschaich/fourfermion}}} both related to the reality of our pseudofermions.
Briefly, the normalization of the gaussian pseudofermion fields $\Phi$ generated at the start of each RHMC trajectory was $\sqrt{2}$ times too large, while the fermion action itself should have involved $\Phi^T \left(\Mdag M\right)^{-1 / 2} \Phi$ rather than $\Phi^T \left(\Mdag M\right)^{-1 / 4} \Phi$, where $M$ is the fermion operator.
These two errors largely counteracted each other, which allowed them to slip past our software tests.\footnote{We thank Jarno Rantaharju for independently checking our results, which helped us track down these problems.}
Following a short review of the lattice theory in \secref{sec:review}, in \secref{sec:results} we present preliminary data obtained with the corrected code, finding broad consistency with our published results.
The main change is a developing signal for spontaneous symmetry breaking in the critical region, potentially consistent with the results of Refs.~\cite{Ayyar:2016lxq, Ayyar:2016nqh}.
We conclude in \secref{sec:conc} with our plans to solidify this result and more carefully study the two transitions that a broken intermediate phase would imply.

\section{\label{sec:review}Review of the lattice theory} 
As described above, we consider four massless reduced staggered fermions in four dimensions.
The lattice action (with sums over repeated indices)
\begin{equation}
  \label{eq:action}
  S = \sum_x \left[\frac{1}{2} \psi^a(x) \eta_{\mu}(x) \De_{\mu}^{ab} \psi^b(x) - \frac{1}{4} G^2 \eps_{abcd} \psi^a(x) \psi^b(x) \psi^c(x) \psi^d(x)\right]
\end{equation}
contains a single-site SU(4)-invariant four-fermion term.
Here $\eta_{\mu}(x) = \left(-1\right)^{\sum_{i < \mu} x_i}$ is the usual staggered fermion phase while $\De_{\mu}^{ab} \psi^b(x) = \frac{1}{2} \de^{ab} \Big\{\psi^b(x + \muhat) - \psi^b(x - \muhat)\Big\}$.
The action possesses exact global symmetries under the transformations
\begin{align}
  \psi(x) & \to e^{i\eps(x) \al} \psi(x)      & & \mbox{with} & \eps(x)      & = \left(-1\right)^{\sum_i x_i} \quad \mbox{and} \quad \al \in \mathrm{su(4)} \cr
  \psi(x) & \to \Ga \psi(x)                   & & \mbox{with} & \Ga          & \in \Big\{1, -1, i\eps(x), -i\eps(x)\Big\}                                    \\
  \psi(x) & \to \xi_{\mu}(x) \psi(x + \muhat) & & \mbox{with} & \xi_{\mu}(x) & = \left(-1\right)^{\sum_{i > \mu} x_i}.                                      \nn
\end{align}
The first of these corresponds to the above-mentioned SU(4) symmetry.
The second is a $Z_4$ subgroup of the usual U(1) symmetry, which is all that the four-fermion interaction preserves.
Finally the staggered shift symmetries in the third line can be considered a discrete remnant of continuum chiral symmetry~\cite{Bock:1992yr}.

Since reduced staggered fermions involve no independent \psibar fields, these symmetries strongly constrain the possible bilinear terms that can arise in the lattice effective action.
Single-site bilinears of the form $\psi^a(x)\psi^b(x)$ violate the SU(4) and $Z_4$ symmetries, while the SU(4)-invariant multilink bilinear operators
\begin{align}
  \label{eq:ops}
  O_1 & = \sum_{x,\,\mu} m_{\mu} \eps(x) \xi_{\mu}(x) \psi^a(x) S_{\mu} \psi^a(x)                                   &
  O_3 & = \sum_{x,\,\mu,\,\nu,\,\la} m_{\mu\nu\la} \xi_{\mu\nu\la}(x) \psi^a(x) S_{\mu} S_{\nu} S_{\la} \psi^a(x)
\end{align}
violate the shift symmetries~\cite{vandenDoel:1983mf, Golterman:1984cy}.
In these expressions
\begin{align*}
  \label{eq:sym}
  \xi_{\mu\nu\la}(x) & \equiv \xi_{\mu}(x) \xi_{\nu}(x + \muhat) \xi_{\la}(x + \muhat + \nuhat) &
  S_{\mu} \psi(x)    & = \psi(x + \muhat) + \psi(x - \muhat)
\end{align*}
and $m_{\mu\nu\la}$ is totally antisymmetric in its indices.
In the absence of explicit symmetry-breaking external sources, bilinear condensates can only form if the corresponding lattice symmetries break spontaneously.

Even in the absence of symmetry-breaking bilinear condensates, a strong-coupling expansion predicts non-zero masses for both fermionic and bosonic excitations.
The leading term in this expansion corresponds to the static $G \to \infty$ limit in which the kinetic operator is dropped.
In the partition function we expand the exponential of the four-fermion term in powers of $G$, obtaining
\begin{equation}
    Z \sim \left[6G^2 \int d\psi^1(x)d\psi^2(x) d\psi^3(x) d\psi^4(x) \psi^1(x)\psi^2(x)\psi^3(x)\psi^4(x)\right]^V
\end{equation}
for lattice volume $V$, which corresponds to saturation by a single-site four-fermion condensate.

A straightforward computation of the fermion propagator $F(x) = \vev{\psi^1(x) \psi^1(0)}$, following the procedure described in \refcite{Eichten:1985ft}, produces the momentum-space expression
\begin{align}
  F(p) & = \frac{i\sqrt{6G^2} \sum_{\mu} \sin p_{\mu}}{\sum_{\mu}\sin^2 p_{\mu} + m_F^2} & & \mbox{with} & m_F^2 & = 4 \left(6G^2\right)^2 - 2.
\end{align}
An analogous calculation for the bosonic propagator $B(x) = \vev{b(x) b(0)}$ of the single-site fermion bilinear $b \equiv \psi^1\psi^2 + \psi^3\psi^4$ leads to
\begin{align}
  B(p) & = \frac{8 \left(6G^2\right)}{4 \sum_{\mu}\sin^2 (p_{\mu} / 2) + m_B^2} & & \mbox{with} & m_B^2 & = 4 \left(6G^2\right) - 8.
\end{align}
In addition to predicting massive fermionic and bosonic excitations at strong coupling, this analysis suggests an interpretation for the symmetric mass generation.
Namely, this may correspond to the condensation of a bilinear formed from the original elementary fermion $\psi^a$ and the composite fermion $\Psi^a \equiv \eps_{abcd} \psi^b \psi^c \psi^d$ that transforms in the complex conjugate representation of the SU(4) symmetry.
The formation of such a four-fermion condensate is clearly a non-perturbative phenomenon invisible in weak-coupling perturbation theory, which motivates our numerical lattice investigations.

As usual, our RHMC calculations use an auxiliary real scalar field $\si_+^{ab} = -\si_+^{ba}$ to bring the action~(\ref{eq:action}) into a form quadratic in the fermion fields,
\begin{equation*}
  S = \sum_x \left[\psi^a(x) \left(\eta.\De^{ab} + G\si_+^{ab}(x)\right) \psi^b(x) + \frac{1}{4}\left(\si_+^{ab}(x)\right)^2\right] = \sum_x \left[\psi^a(x) M^{ab}(x) \psi^b(x) + \frac{1}{4}\left(\si_+^{ab}(x)\right)^2\right],
\end{equation*}
defining the fermion operator $M^{ab}(x) \equiv \eta.\De^{ab} + G\si_+^{ab}(x)$.
The subscript indicates that $\si_+^{ab}$ is self-dual,
\begin{equation}
  \label{eq:proj}
  \si_+^{ab} = \frac{1}{2}\left(\si^{ab} + \frac{1}{2} \eps_{abcd} \si^{cd}\right) \equiv P_+^{abcd} \si^{cd}.
\end{equation}
Now we can integrate over the fermions to produce the pfaffian $\pf{M(\si_+)}$, which turns out to be positive semi-definite.
This follows from the fact that $M$ is real, anti-symmetric, and invariant under one of the SU(2) subgroups of the $\mathrm{SO(4)} \simeq \mathrm{SU(2)} \times \mathrm{SU(2)}$ global symmetry of the quadratic action above.
The eigenvalues of $M$ are therefore pure imaginary and come in pairs $i\la$ and $-i\la$, both of which are doubly degenerate due to the invariance under this SU(2).
We have checked this conclusion numerically.
It forbids sign changes in the pfaffian, which would correspond to an odd number of eigenvalues passing through zero as $\si_+$ varies.

\section{\label{sec:results}Numerical results for the phase diagram} 
In order to directly search for spontaneous symmetry breaking we have augmented the lattice action~(\ref{eq:action}) by adding three source terms,
\begin{equation}
  \label{eq:sources}
  \De S = \sum_{x,\,a,\,b} \Big(m_1 + \eps(x)m_2\Big) \left[\psi^a(x) \psi^b(x)\right]_+ \Si^{ab} + m_3\sum_{x,\,\mu,\,a} \eps(x) \xi_{\mu}(x) \psi^a(x) S_{\mu} \psi^a(x).
\end{equation}
As in \eq{eq:proj} we consider the self-dual part of the single-site bilinear in the first set of terms, which we couple to the SU(4)-breaking source $\displaystyle \Si^{ab} = \left(\begin{array}{cc} i\si_2 & 0 \\ 0 & i\si_2\end{array}\right)$.
In this proceedings we work with $m_1 = 0$, considering only the ``staggered'' single-site bilinear with coupling $m_2$.
The latter operator breaks all the exact symmetries of the action but appears as a rather natural mass term when the model is rewritten in terms of two full staggered fields.
The final term corresponds to the shift-symmetry-breaking one-link operator $O_1$ in \eq{eq:ops}.

\begin{figure}[btp]
  \includegraphics[width=0.48\linewidth]{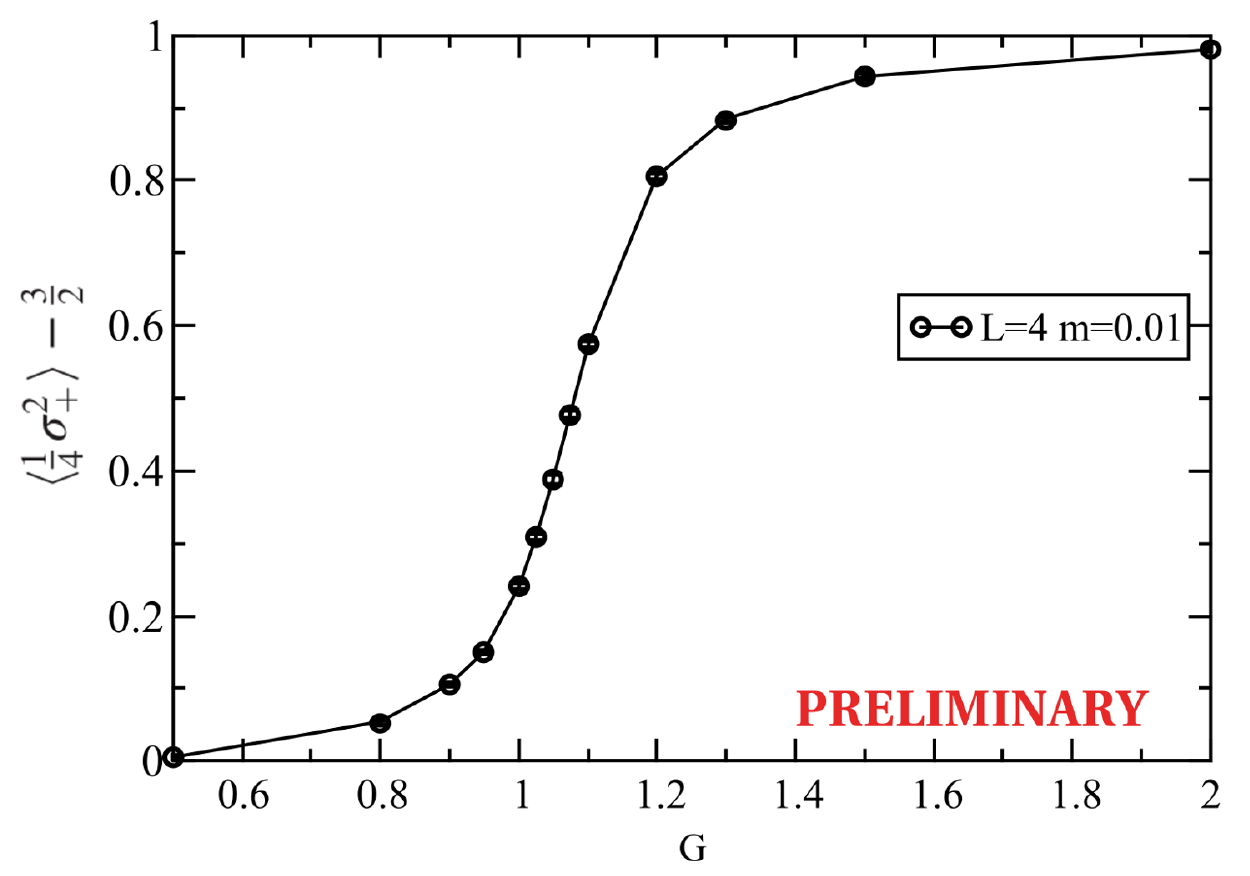} \hfill \includegraphics[width=0.48\linewidth]{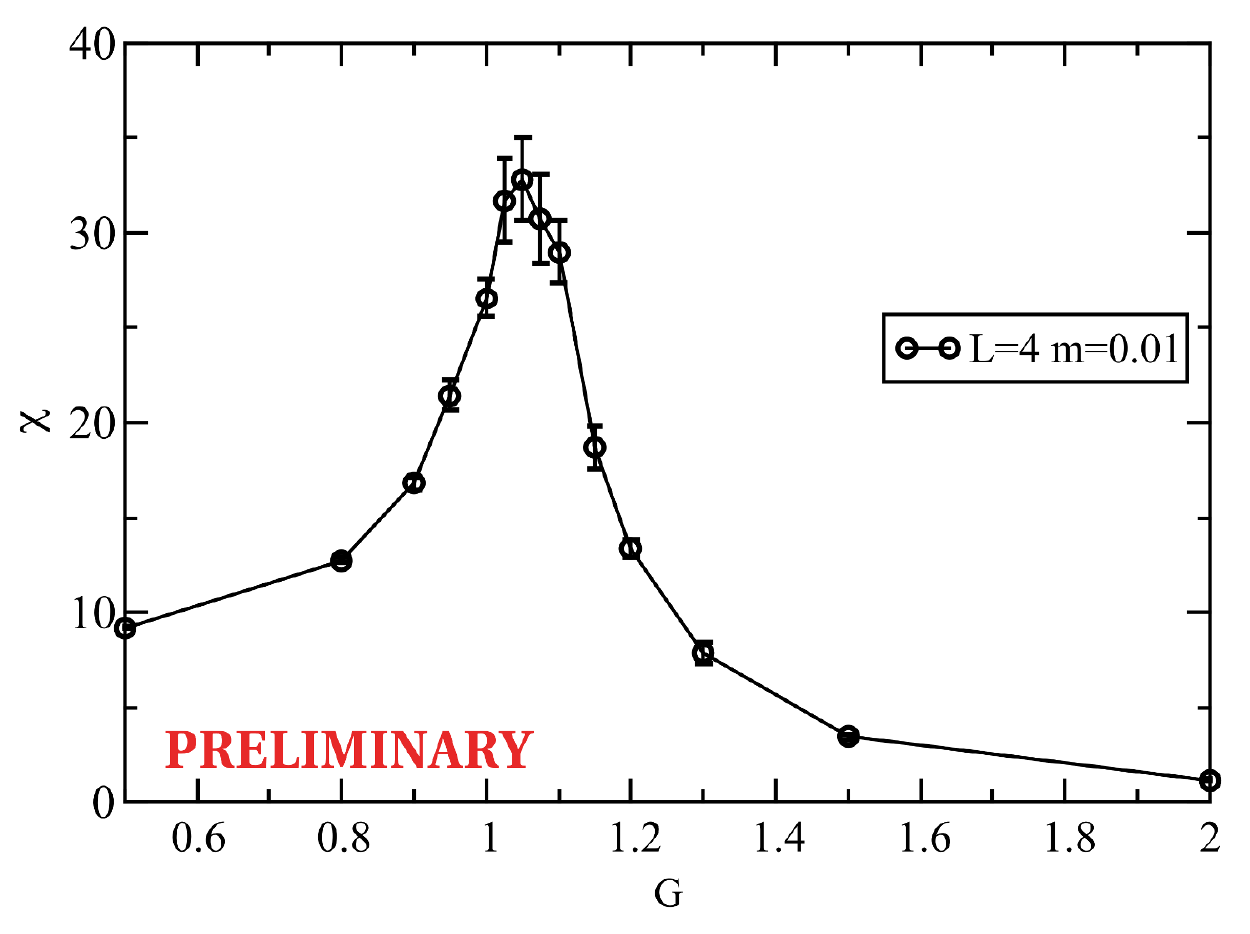}
  \caption{\label{fig:transition}\textbf{Left:} $\frac{1}{4} \vev{\si_+^2} - \frac{3}{2}$ serves as a proxy for the four-fermion condensate, increasing from zero at weak four-fermion coupling $G$ to unity at strong coupling.  \textbf{Right:} The staggered susceptibility $\chi$ peaks in a narrow critical region $1 \lsim G \lsim 1.1$.  Both are measured on $4^4$ lattices with $m_2 = 0.01$ while $m_1 = m_3 = 0$.}
\end{figure}

As mentioned in the introduction, our corrected numerical results are broadly consistent with those published in \refcite{Catterall:2016dzf}.
\fig{fig:transition}, for example, reproduces the behavior shown in Figs.~1 and 14 of \refcite{Catterall:2016dzf} for two observables that are sensitive to the transition from weakly coupled free fields to strongly coupled four-fermion condensates.
The square of the auxiliary field $\frac{1}{4} \si_+^2 = \frac{1}{2} \sum_{a < b}\left(\si_+^{ab}\right)^2$ in the left plot serves as a proxy for the four-fermion condensate.
A simple analytic calculation predicts $\frac{1}{4} \vev{\si_+^2} \to \frac{3}{2}$ as $G \to 0$ and $\frac{1}{4} \vev{\si_+^2} \to \frac{5}{2}$ as $G \to \infty$, which our numerical results reproduce.
We see a continuous interpolation between these two limits, but this is not significant given the small $4^4$ lattice volume and non-zero $m_2 = 0.01$ considered so far.

The right plot of \fig{fig:transition} shows the staggered susceptibility
\begin{align}
  \chi & = \frac{1}{V}\left(\vev{O_{\text{stag}}^2} - \vev{O_{\text{stag}}}^2\right) &
  O_{\text{stag}} & = \sum_x \eps(x) \left[\psi^0(x) \psi^1(x)\right]_+
\end{align}
computed on these same $4^4$ ensembles.
As in Fig.~14 of \refcite{Catterall:2016dzf} we see a clear peak in the susceptibility centered around $G_c \approx 1.05$.
The height of this peak is significantly larger than we saw before, but this is related to the non-zero $m_2 = 0.01$ we use in \fig{fig:transition}.
Both plots in this figure indicate a narrow critical region in the same range $1 \lsim G \lsim 1.1$ that we observed previously.

\begin{figure}[btp]
  \centering
  \includegraphics[width=0.5\linewidth]{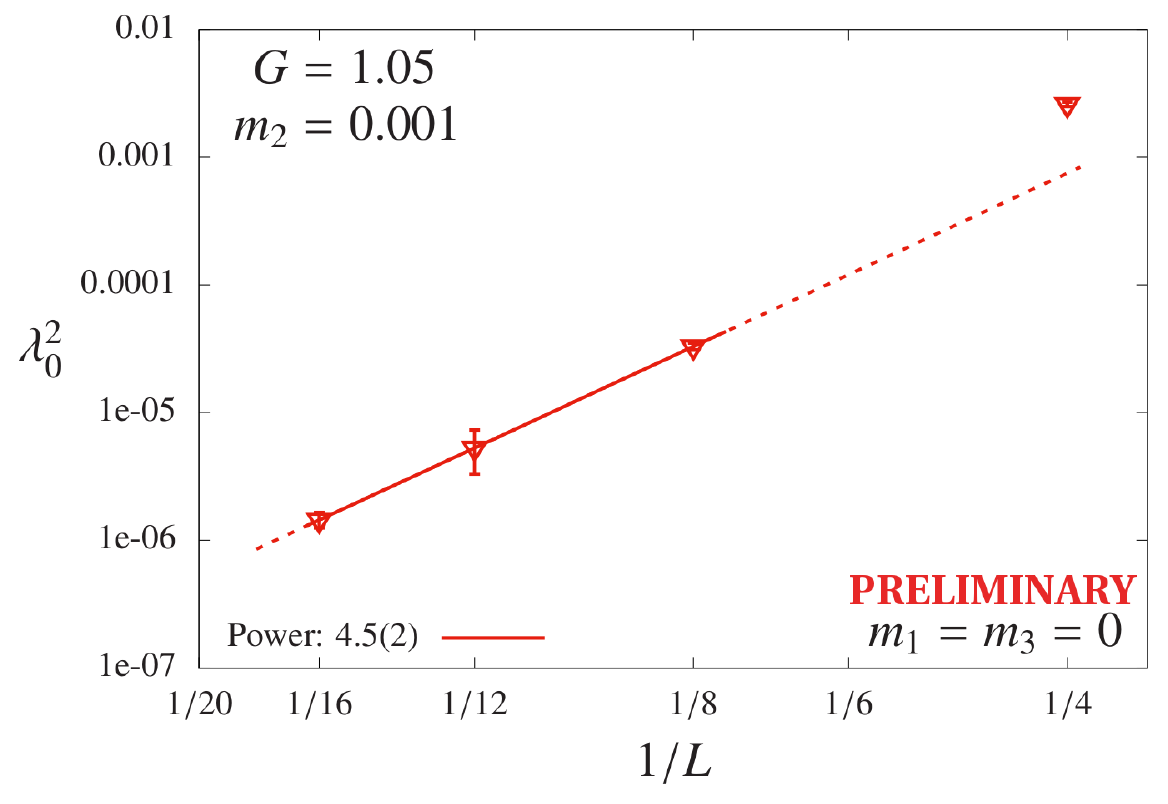}
  \caption{\label{fig:eig}The smallest (squared) eigenvalue of the fermion operator vs.\ $1 / L$ on log--log axes, for fixed $G = 1.05$ and $m_2 = 0.001$ with $m_1 = m_3 = 0$.  A power-law fit to the three largest volumes produces $\la_0^2 \propto L^{-9 / 2}$.} 
\end{figure}

We are currently scanning in $G$ for larger volumes with all three $m_i = 0$, which we expect will reproduce the volume scaling of the peak susceptibility ($\chi_{\text{peak}} \sim L^4$) reported by \refcite{Ayyar:2016lxq}.
In \refcite{Catterall:2016dzf} we identified two other signals of long-range correlations in this critical region: the mass of a composite boson was very small throughout $1 \leq G \leq 1.1$ while the smallest eigenvalue of the fermion operator decreased rapidly with the volume in this regime.
While we have not yet repeated our bosonic two-point function analyses, \fig{fig:eig} shows that the smallest eigenvalue at $G = 1.05$ continues to decrease rapidly, $\la_0 \propto L^{-9 / 4}$.
We use a small but still non-zero $m_2 = 0.001$, which significantly increases $\la_0$ compared to the $m_2 = 0$ results shown in Fig.~5 of \refcite{Catterall:2016dzf}.
The corresponding reduction in critical slowing down has allowed us to investigate larger $L = 16$ than we could easily reach with $m_2 = 0$.

Finally we revisit our search for spontaneous symmetry breaking in the critical region that may correspond to a narrow intermediate phase.
Fixing $G = 1.05$ (and $m_1 = 0$), we compute the vacuum expectation values of the staggered site and one-link bilinears as functions of $m_2$, $m_3$ and the lattice volume.
The presence of the source terms leads to non-zero vevs for the corresponding operators, which on a fixed lattice volume must vanish as the couplings are sent to zero, due to the exact lattice symmetries that appear in that limit.
A signal of spontaneous symmetry breaking would be a condensate that grows with the volume for small values of the external source, which would allow for the possibility that the condensate remains finite in the thermodynamic limit as the source is removed.

\begin{figure}[btp]
  \includegraphics[width=0.48\linewidth]{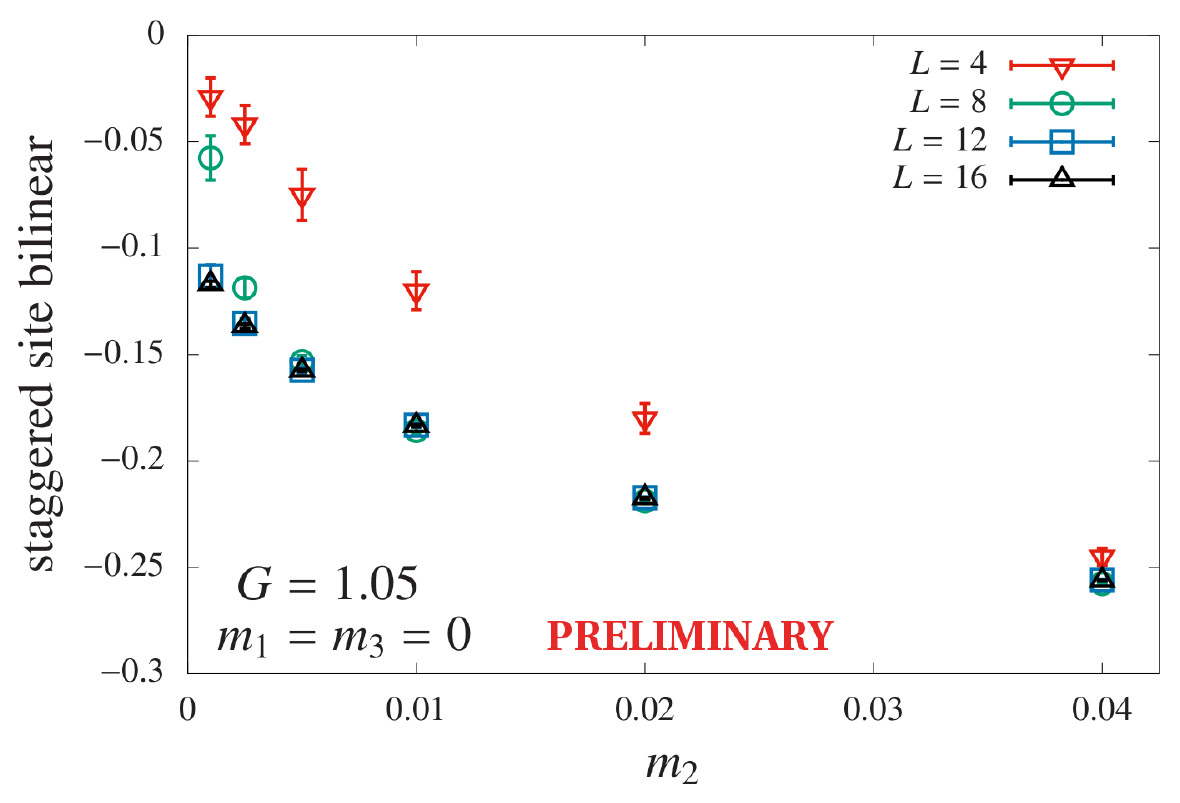} \hfill \includegraphics[width=0.48\linewidth]{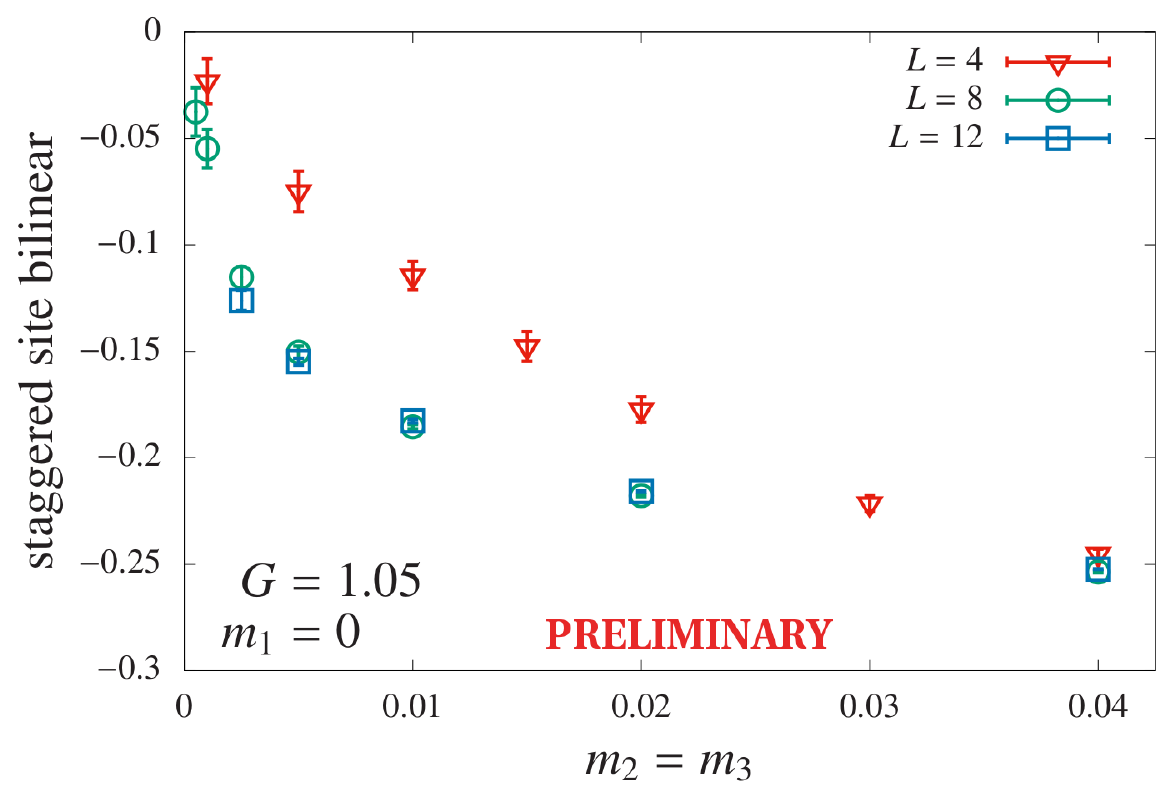}
  \caption{\label{fig:bi}The staggered site bilinear vs.\ $m_2$ for $G = 1.05$ and $m_1 = 0$.  In the left plot we consider $L = 4$, 8, 12 and 16 with no source for the one-link bilinear, $m_3 = 0$.  In the right plot we set $m_3 = m_2$ and study $L = 4$, 8 and 12 as in \protect\refcite{Catterall:2016dzf}.}
\end{figure}

In \fig{fig:bi} we see initial signs of such behavior in the staggered site bilinear.
The two plots in the figure differ only in the value of the one-link coupling: $m_3 = 0$ in the left plot while the right plot has $m_3 = m_2$ as in \refcite{Catterall:2016dzf}.
The staggered site bilinear shows no visible dependence on $m_3$, with results for $4 \leq L \leq 16$ following a single curve for large $m_2 \gsim 0.04$.
In both plots we can see the smallest-volume $4^4$ results departing from this curve and moving towards zero for $m_2 \lsim 0.03$, while the next $8^4$ results follow suit around $m_2 \lsim 0.003$.
However, for $m_3 = 0$ in the left plot we have begun investigating larger $L = 16$, and find that the corresponding staggered site bilinear is no larger than the $L = 12$ results for all $m_2 \geq 0.001$ we have reached so far.
While the small-volume behavior in \fig{fig:bi} supports spontaneous SU(4) symmetry breaking at $G = 1.05$, we hope to strengthen this conclusion with more data on larger volumes in the near future.

\begin{figure}[btp]
  \centering
  \includegraphics[width=0.5\linewidth]{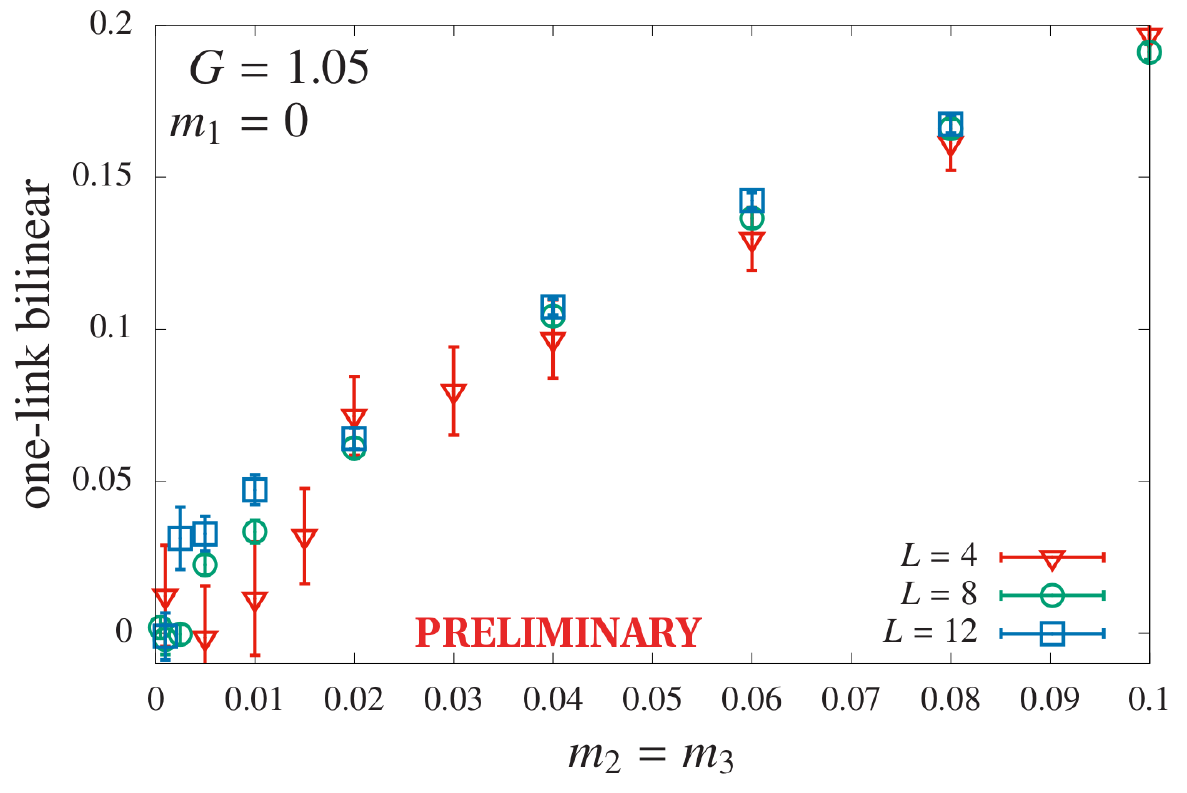}
  \caption{\label{fig:link}The one-link bilinear vs.\ $m_3 = m_2$ for $L = 4$, 8 and 12 at $G = 1.05$ and $m_1 = 0$.}
\end{figure}

\fig{fig:link} carries out the same exercise for the one-link bilinear, using the same $m_3 = m_2$ ensembles considered in the right plot of \fig{fig:bi}.
(When $m_3 = 0$ the one-link condensate vanishes.)
Here we do not see any sign that the staggered shift symmetries break spontaneously.
The results are qualitatively the same as those in Fig.~12 of \refcite{Catterall:2016dzf}, with no significant volume dependence visible for any $m_3 \leq 0.1$.

\section{\label{sec:conc}Conclusions and next steps} 
While the system we study is perhaps the simplest four-fermion theory in four dimensions, it exhibits interesting behavior that motivates further work, both theoretical and computational.
With our corrected code we observed the same large-scale phase structure as in \refcite{Catterall:2016dzf}, with a narrow critical region around $1 \lsim G \lsim 1.1$ separating the symmetric massless weak-coupling (PMW) phase from the symmetric massive strong-coupling (PMS) phase.
We continue to see signs of long-range correlations in this critical region, and have now begun to see a developing signal consistent with spontaneous SU(4) symmetry breaking at $G = 1.05$.

We are already working to improve \fig{fig:transition} by repeating our scans in the four-fermion coupling $G$ on larger volumes with zero external sources.
This will provide another look at the finite-size scaling of the peak in the staggered susceptibility, and hopefully will allow us to resolve the two transitions that a spontaneously broken intermediate phase would imply.
Further studies would then be needed to distinguish whether the transitions are continuous or first order, and to estimate critical exponents in the former case, with the goal of exploring whether they may provide the possibility of a new continuum limit for strongly interacting fermions.

It is also important to improve the scans in $m_2$ shown in \fig{fig:bi}, to strengthen the developing signs of spontaneous symmetry breaking at $G = 1.05$.
Repeating this exercise in the PMW and PMS phases (e.g., for $G \approx 0.85$ and 1.25, respectively) should provide useful contrasts that may clarify the nature of the critical regime.
It seems likely we will need to consider smaller $m_2 \ll 10^{-3}$ at $G = 1.05$, which will be challenging due to the critical slowing down we observe in the transition region when $m_2 = 0$.
The rapid decrease in the smallest eigenvalues of the fermion operator corresponds to an increasing condition number in each multi-mass conjugate gradient inversion of $\Mdag M$, which dominates the computational cost of the RHMC algorithm.
It may be necessary to implement significant extensions to the code to make these calculations practical, for example applying staggered deflation or multigrid techniques.

Independently of these numerical investigations, there is also work ongoing to explore the conceptual issues connected to the observed behavior.
For example, it has been proposed~\cite{You:2014vea, BenTov:2015gra, Wang:2013yta} that similar quartic interactions can be used in the context of domain wall fermions to realize a lattice regularization of chiral gauge theories along the lines originally proposed by \refcite{Eichten:1985ft}. 
Although these proposals describe non-relativistic fermions using hamiltonian language, it is nevertheless intriguing that the sixteen Majorana fermions they require match the sixteen Majorana fermions that are expected at weak coupling in this lattice theory.
However, it is not clear to the authors whether these proposals circumvent all the difficulties described in \refcite{Poppitz:2010at}.
Even the possibility of a new strongly interacting continuum limit is subtle.
The reduced staggered fermions used to define the lattice theory transform under a twisted group comprising both Lorentz and flavor symmetries~\cite{Banks:1982iq}, suggesting that even if a new fixed point exists it might not be Lorentz invariant.

More recently there have been interesting proposals concerning the novel phase structures of the three- and four-dimensional theories.
In three dimensions, for example, the apparent direct transition between the weakly and strongly coupled symmetric phases might be understood within the theoretical framework of deconfined quantum criticality~\cite{You:2017ltx}.
In four dimensions \refcite{Catterall:2017ogi} argued that the fermion mass in the PMS phase may result from the proliferation of topologically non-trivial defects with non-zero Hopf invariant, following a similar logic to Witten's treatment of the two-dimensional Thirring model~\cite{Witten:1978qu}.
We look forward to further work on this subject, in particular continuing interplay between non-perturbative numerical investigations and conceptual considerations.

\vspace{10 pt}
\noindent {\sc Acknowledgments:}~We thank Shailesh Chandrasekharan and Jarno Rantaharju for useful discussions. 
This work is supported in part by the U.S.~Department of Energy, Office of Science (DOE), Office of High Energy Physics, under Award Number DE-SC0009998. 
Numerical calculations were carried out on the DOE-funded USQCD facilities at JLab.

\raggedright
\bibliography{Lattice2017_109_SCHAICH}
\end{document}